\documentclass[a4paper,11pt]{article}
\usepackage{jinstpub} 
\usepackage{lineno}
\usepackage{tikz}
\usepackage{graphicx}

\title{\boldmath Beam test of n-type Silicon pad array detector at PS CERN}

\author[a]{Sawan}
\author[b]{M.~Bregant}
\author[c]{J.~L.~Bouly}
\author[c]{O.~Bourrion}
\author[d]{A.~van den Brink}
\author[e]{T.~Chujo}
\author[b1]{C.~Krug}
\author[f]{L.~Kumar}
\author[a]{V.~K.~S.~Kashyap}
\author[g]{A.~Ghimouz}
\author[h]{M.~Inaba}
\author[i]{T.~Isidori}
\author[j]{C.~Loizides}
\author[a,1]{B.~Mohanty\note{Corresponding author.}}
\author[a]{M.M.~Mondal}
\author[i]{N.~Minafra}
\author[j]{N.~Novitzky}
\author[b]{N.~Ponchant}
\author[k]{M.~Rauch}
\author[a]{K.~P.~Sharma}
\author[a]{R.~Singh}
\author[l]{D.~Thienpont}
\author[c]{D.~Tourres}
\author[a]{G.~Tambave}

\affiliation[a]{National Institute of Science Education and Research, Homi Bhabha National Institute, Jatni, India}
\affiliation[b]{Universidade de São Paulo (USP), São Paulo, Brazil}
\affiliation[c]{Universit\'e Grenobles Alpes, CNRS, Grenoble, France}
\affiliation[d]{Institute for Gravitational and Subatomic Physics (GRASP), Utrecht University, Netherlands}
\affiliation[e]{ Institute of Pure and Applied Sciences, University of Tsukuba, Tsukuba, Japan}
\affiliation[b1]{Instituto de Física, Universidade Federal do Rio Grande do Sul, Porto Alegre, RS, Brazil}
\affiliation[f]{Physics Department, Panjab University, Chandigarh, India}
\affiliation[g]{Paul Scherrer Institut, Switzerland}
\affiliation[h]{Tsukuba University of Technology, Japan}
\affiliation[i]{The University of Kansas, Lawrence, USA}
\affiliation[j]{ORNL, Oak Ridge, USA}
\affiliation[k]{University of Bergen, Bergen, Norway}
\affiliation[l]{Ecole Polytechnique, CNRS/IN2P3, Omega, Palaiseau, France}

\emailAdd{bedanga@niser.ac.in}

\abstract{This work reports the testing of a Forward Calorimeter (FoCal) prototype based on an n-type Si pad array detector at the CERN PS accelerator. The FoCal is a proposed upgrade in the ALICE detector operating within the pseudorapidity range of 3.2 < $\mathrm{\eta}$ < 5.8. It aims to measure direct photons, neutral hadrons, vector mesons, and jets for the study of gluon saturation effects in the unexplored region of low momentum fraction x ($\mathrm{\sim10^{-5} - 10^{-6}}$). The prototype is a $\mathrm{8\times9}$ n-type Si pad array detector with each pad occupying one cm$^2$ area, fabricated on a 6-in, 325~$\mathrm{\pm 10 \thinspace \mu}$m thick, and high-resistivity ($\sim$7 k$\Omega \thinspace$ cm) Si wafer which is readout using HGCROCv2 chip. The detector is tested using pion beams of energy 10~GeV and electron beams of energy 1-5~GeV. The measurements of the Minimum Ionizing Particle (MIP) response of pions and the shower profiles of electrons are reported.}


\begin{document}
\maketitle
\flushbottom

\section{Introduction}
\label{sec:intro}
Calorimeters are essential instruments in particle and nuclear physics, designed to measure the energy of particles by fully absorbing them and converting their energy into a detectable form. They play a fundamental role in experiments, distinguishing between different types of particles through their interactions with detector materials. There are two main types of calorimeters: electromagnetic and hadronic. Electromagnetic calorimeters detect electrons, positrons, and photons by their electromagnetic interactions. Hadronic calorimeters detect hadrons by their strong and electromagnetic interactions. Some electromagnetic calorimeters have a segmented sampling design with layers of passive (absorber) and active (detector) materials. This allows for precise measurements of the position and energy of particles in the longitudinal and transverse directions. A detector named Forward Calorimeter (FoCal), composed of a segmented sampling calorimeter and a hadronic calorimeter, is proposed for the ALICE detector upgrade during the forthcoming Long Shutdown 3, bringing new opportunities for data collection and exploration in LHC Run 4~\cite{focal_LOI}. FoCal aims to explore QCD in the low-x~($\mathrm{\sim10^{-5} - 10^{-6}}$) regime via the measurement of direct photons, neutral hadrons, vector mesons, and jets at forward rapidity, where the high gluon density produces non-linear effects such as gluon saturation~\cite{physics_of_focal}.

The electromagnetic portion of FoCal~(FoCal-E) is a segmented calorimeter alternating a total of 20 tungsten absorber layers (3.5~mm-thick corresponding to 1 radiation length $\mathrm{X_0}$) with 18 low-granularity (pad size 1~$\times$~1~cm$^{2}$) and two high-granularity (pixel size 30~$\times~$30~$\mathrm{\mu m^2}$) sensor layers. The low-granularity layers enable the measurement of the total energy and position of the particles, and the high-granularity layers are used for separating the shower of direct photons from the background, mainly from the decay of neutral pion and eta mesons. In the past, several prototypes of the Si-W electromagnetic calorimeter have been built and tested using different geometries and segmentation of the Si sensors. These include the high-granularity based pixel sensors~\cite{pixel_paper1, Pixel_paper2_calice}, the low-granularity based pad sensors with individual pad size of 1$\times$1 cm$^2$~\cite{sanjib_paper, Clice_pads1, Clice_pads2, Tsukuba_paper_pads} and the large p-type Si sensors developed on 6-inch wafer~\cite{japanese_sps_pads}. This paper discusses the new n-type $\mathrm{8\times9}$ Si sensors fabricated on a 6-in wafer, which are read using a highly integrating HGCROCv2 chip \cite{HGCROCv2_paper}.

The subsequent sections will discuss the calorimeter design and readout electronics (Sec.~\ref{sec: calorimeter designs and readout electronics}), the setup of the detector with the trigger in the test beam (Sec.~\ref{sec: test beam setup}), the estimation of the background noise (Sec.~\ref{sec: background estimation}) and the characterization of the detector's response to pion and electron beams (Sec.~\ref{sec: response to pion beam} \& \ref{sec: response to electron beam}) spanning different energy ranges.


\section{Calorimeter design and Readout electronics} \label{sec: calorimeter designs and readout electronics}

\begin{figure}[t]
\centering
\rotatebox{270}{\includegraphics[width=.6\textwidth]{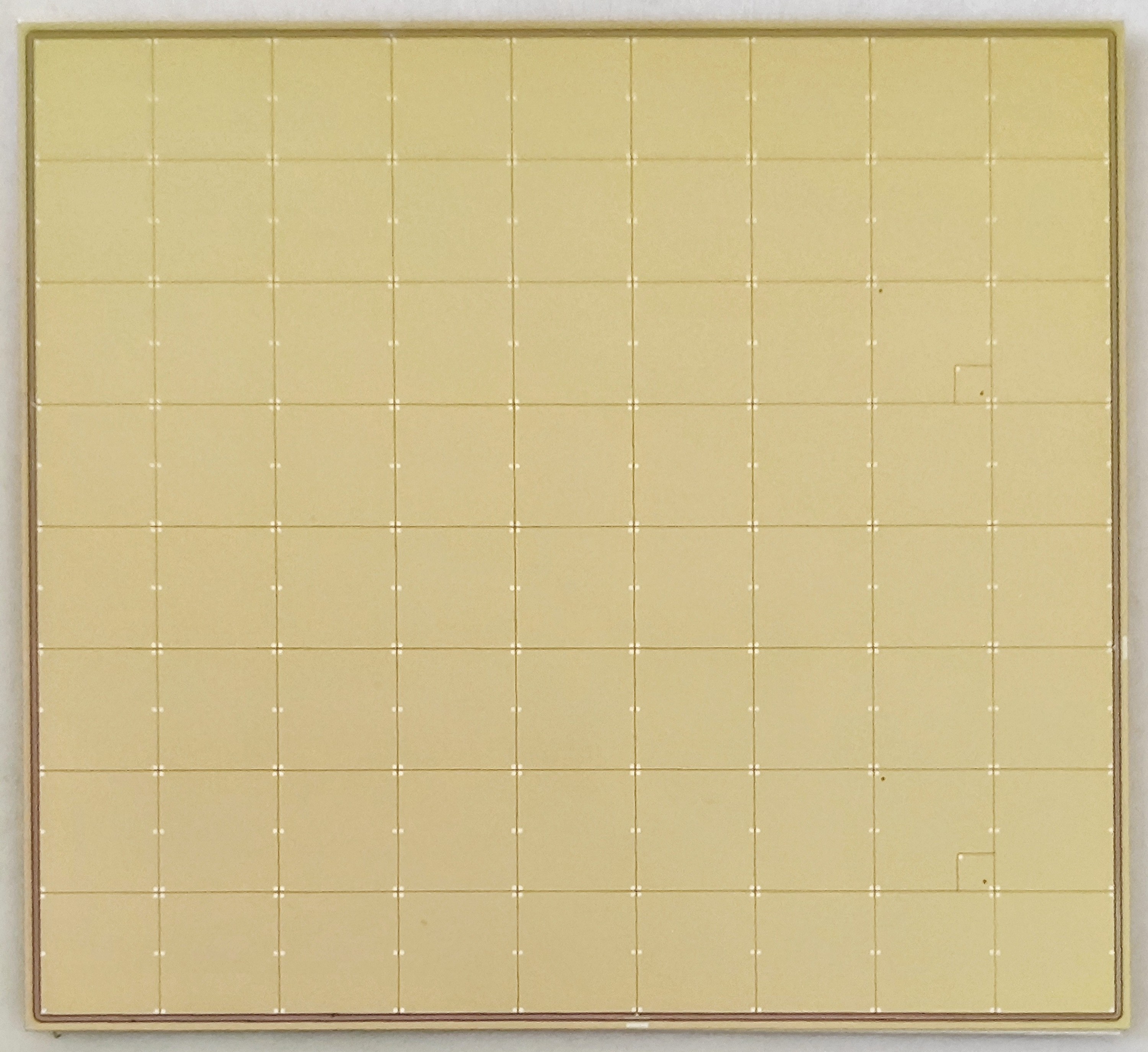}}
\caption{$\mathrm{8\times9}$ Si pad array diced from a 6-in Si wafer.}\label{fig:wafer_image}
\end{figure}

 \subsection{Silicon pad array}
 The silicon pad array is fabricated at Bharat Electronics Limited, Bangalore \cite{bel} from an n-type 6-in silicon wafer with a thickness of 325~$\pm$~10~$\mathrm{\mu m}$ and high resistivity ($\sim$7~k$\Omega \thinspace$~cm). The choice of high resistivity is essential to minimize the dark current or the leakage current in the detector. From the 6-in Si wafer, a precise array of 72 pads is created in a $\mathrm{8\times9}$ configuration, with each individual pad covering an area of 1~cm$^{2}$. The diced wafer is placed in a jig to ensure proper alignment and attachment with the \textit{Printed Circuit Board} (PCB) \cite{oliver_paper}. The sensor is glued to the PCB using SADER epoxy glue and kept overnight for the glue to settle down. The connection between the wafer and PCB is made with gold wire bonds.
 
Each pad in the pad array is designed with specific coatings for optimal performance. On the p-side (wire bond side), a thin layer of aluminum (Al) metal followed by silicon nitride is applied. This protective coating shields the pad from ambient light and moisture. Each pad's n-side (glue side) is coated with a maximum thickness of 6~$\mathrm{\mu m}$ of aluminum, forming an Ohmic contact. This configuration ensures efficient charge transfer between the metal contact and the semiconductor. 
The silicon pad array also incorporates four \textit{Guard Rings} (GR); the innermost GR is kept at reverse-bias voltage while the middle two are floating, and the outermost is connected to the ground. The guard rings are used to protect the pad cells from the breakdown.

\subsection{Readout electronics}
The readout electronics for the n-type Si pad array detector consist of a \textit{High Granularity Calorimeter ReadOut Chip} (HGCROCv2)~\cite{HGCROCv2_paper} hosted by the PCB. This chip is being developed for the \textit{High Granularity Calorimeter} (HGCal) upgrade project of the CMS experiment at CERN. 
The HGCROCv2 consists of 72 channels of a pre-amplifier, a shaping amplifier, and a 10-bit ADC operated at 40~MHz clock with selectable dynamic range (80~fC, 160~fC, and 320~fC). Once the ADC saturates, the charge up to 10~pC can be acquired using a time-to-digital converter (TDC) based on the Time-over-Threshold (TOT) method. In addition, the TDC is used to record the time of arrival (TOA). There is a DRAM memory in place to keep the charge and timing information for 512 \textit{Bunch Crossings} (BC) occurring at a rate of 40 MHz. Once a trigger signal is received, the data is read out and processed. Two dedicated 1.28~Gbps high-speed links are used to ship the charge and the time information to the KCU105 \textit{Data acquisition} (DAQ) board~\cite{KCU105}. A dedicated firmware and software are developed to control, configure, and monitor the HGCROCv2 ASIC. A detailed explanation of the data acquisition system is published in the research article~\cite{oliver_paper}.  

\section{Test beam detector setup at PS CERN} \label{sec: test beam setup}
\subsection{Beam characteristics in PS T9 area}
A dedicated test setup was commissioned in the T9 area at \textit{Proton Synchrotron} PS CERN to test the detector. The T9 area is an experimental facility in the east area of the PS accelerator. It uses a secondary beam of particles that is produced by the impact of a primary proton beam with a target. The primary proton beam has a momentum of 24~GeV$/c$. The beam is operated in two different modes - mixed hadron mode and the electron mode. The mixed hadron mode uses a 200~mm Al target (hadron-enriched) to produce a variety of particles, such as electrons, muons, and hadrons, with momenta from 0.5~GeV$/c$ to 16~GeV$/c$. The purity of the pion beam for momenta above 5~GeV$/c$ is close to 100\%. The electron mode uses either a 200~mm Be target or a 3~mm W target to generate electrons with momenta up to 5~GeV$/c$. The electron beam has more than 90\% purity for momenta below 5~GeV$/c$ and reaches up to 99\% purity at 1~GeV$/c$. A Cherenkov counter is used to filter out the pure electrons further. In this experiment, the pion beams with energy of 10~GeV are used, while electrons with energy from 1-5~GeV are selected using the Cherenkov detector.


\subsection{Detector setup and trigger scheme}
\begin{figure}[t]
\centering
\includegraphics[width=.6\textwidth]{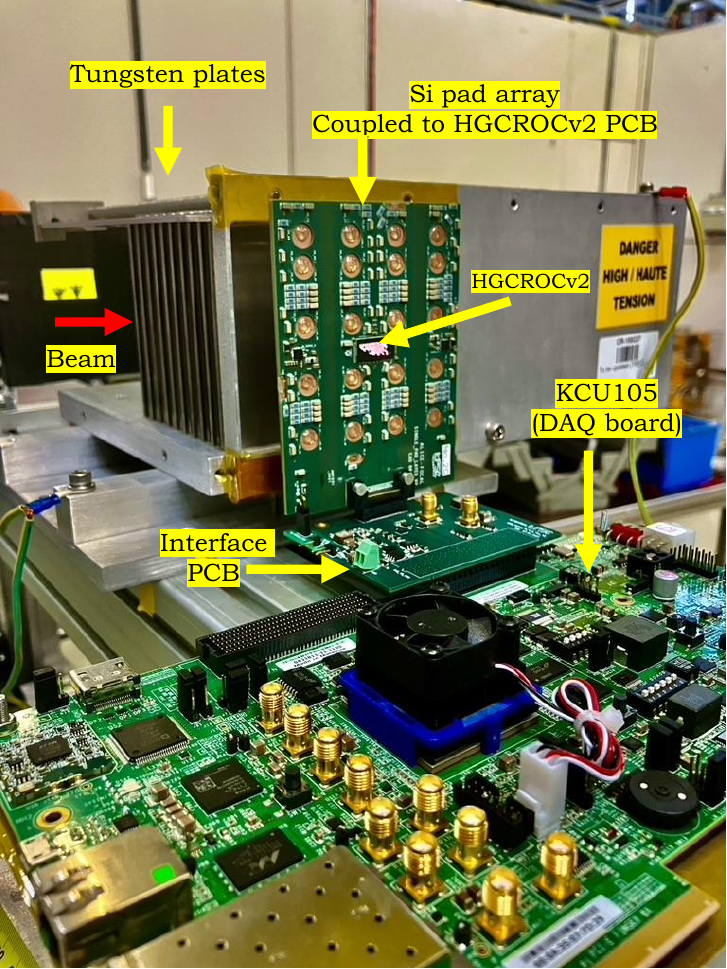}
\caption{Test setup showing the Si pad array, its readout electronics, and the tungsten plates.\label{fig:test_setup}}
\end{figure}

The detector assembly consists of a silicon pad array + PCB + HGCROCv2, connected to the DAQ board through an interface board. A source meter for the constant supply of biasing voltage to the detector. An aluminum mechanical structure that holds the 3.5~mm thick tungsten absorbers with a separation of 5~mm between them, as shown in Fig.~\ref{fig:test_setup}. The absorbers can be added or removed from the structure to obtain the electron shower at different thicknesses.
\begin{figure}[t]
\centering
\includegraphics[width=0.9\textwidth]{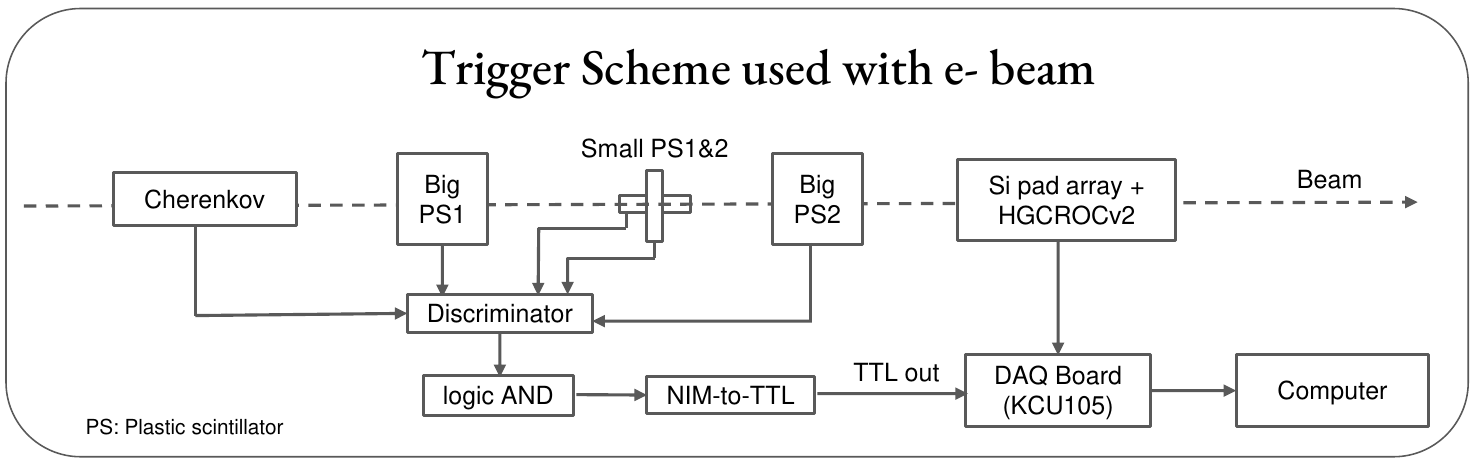}
\caption{Trigger schematic used for electron beam. The Cherenkov signal was not included in the trigger logic during the measurements done with the pion beam as the beam purity was close to 100$\%$ within the studied energy range.\label{fig:trigger-scheme}}
\end{figure}
The experiment uses four plastic scintillators and a Cherenkov counter to generate triggers. Two large plastic scintillators with an area of 13$\times$11~cm$^2$ cover the geometrical acceptance of the detector. Two small plastic scintillators, which overlap in a cross shape with an area of 1.4~cm$^{2}$, provide a trigger to select the particles with an incidence around a pad.
The AND logic of these four plastic scintillators, along with the Cherenkov (used for electrons), is fed to the discriminator, which then converts the signal from \textit{Nuclear Instrument Modules} (NIM) to \textit{Transistor-Transistor Logic} (TTL), which is compatible with the DAQ board. Finally, the DAQ board, upon receiving the trigger signal, transmits the data to the computer for further analysis. The trigger schematic is shown in the Fig.~\ref{fig:trigger-scheme}.

\section{Results and Interpretation}
\begin{figure}[htbp]
\centering
\includegraphics[width=0.9\textwidth]{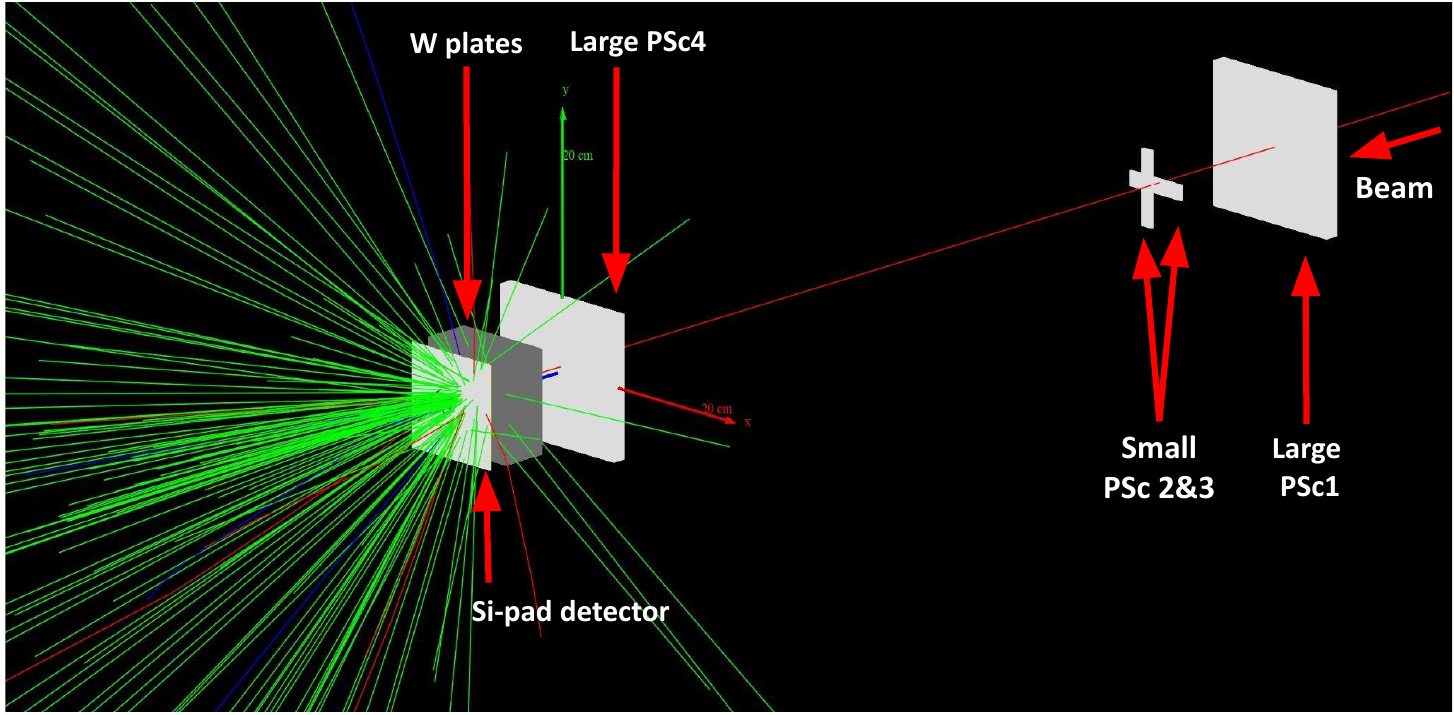}
\caption{Test beam setup modeled in Geant4 simulation framework}~\cite{geant4package}.\label{fig:sim_setup}
\end{figure}

The first part of the section reports the noise estimation and the test results obtained using a pion beam focused on different Si pad array cells. The second part reports the test beam results for the electron data obtained by placing tungsten plates in front of the detector.
These results are also compared with the Geant4 simulations~\cite{geant4package}. The test beam geometry constructed in Geant4 is shown in Fig.~\ref{fig:sim_setup}, where the coordinate system is defined by the position of the large plastic scintillator PSc4, which is placed on the origin. The x and y axes are indicated by green and red arrows, respectively, with the green arrow pointing upwards and the red arrow pointing to the left as seen from the beam direction, which travels from right to left. The simulations use the same detector material and composition as discussed in the paper~\cite{japanese_sps_pads}, except for the absence of a Hadronic calorimeter and the silicon sensors in the gap between W plates. A silicon sensor is placed at a distance of 15mm from the last W plate. The number of W plates also varies in the simulations according to the test setup for obtaining the electromagnetic shower at different depths.

\subsection{Background noise estimation} \label{sec: background estimation}
\begin{figure}[t]
\centering
\includegraphics[width=0.47\textwidth]{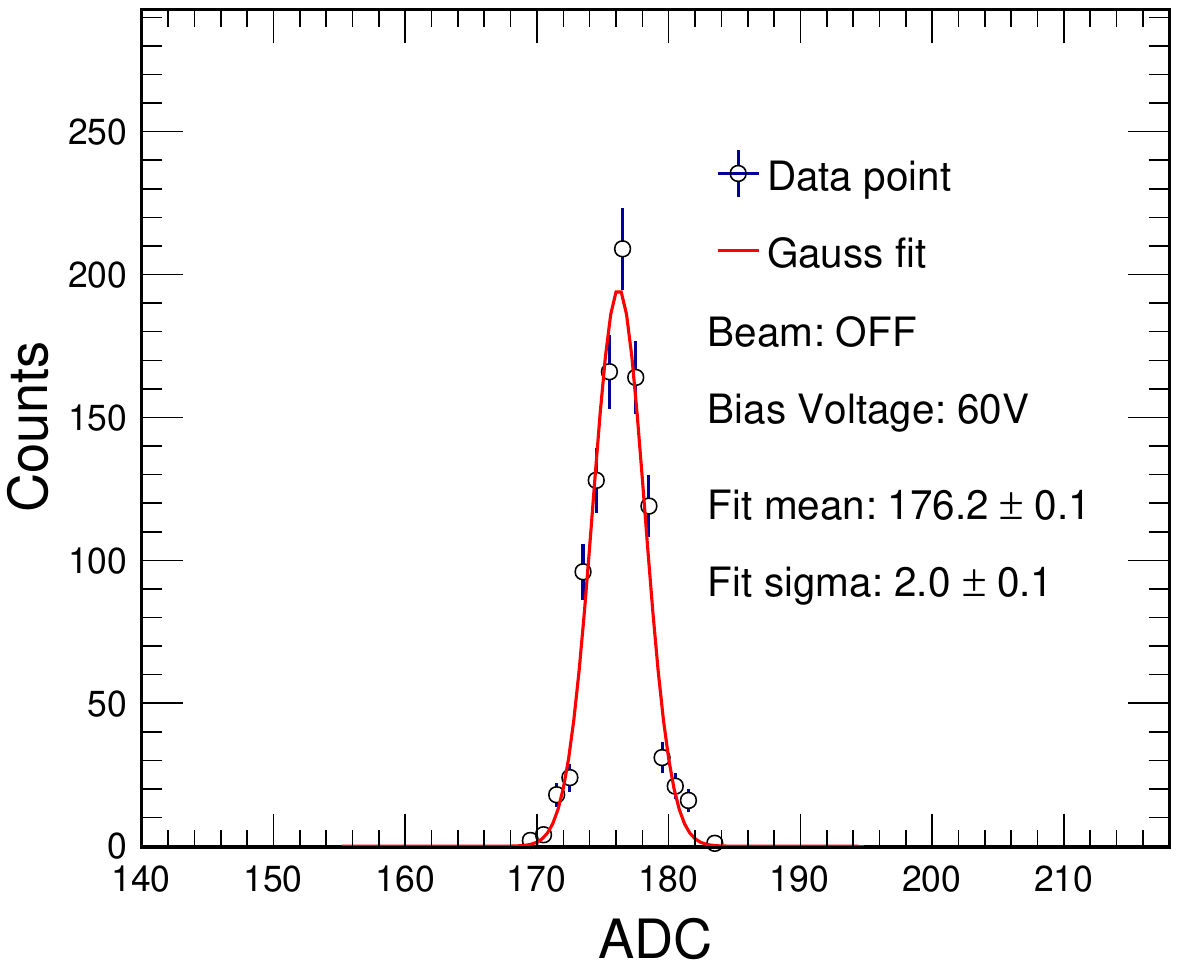}
\qquad
\includegraphics[width=0.47\textwidth]{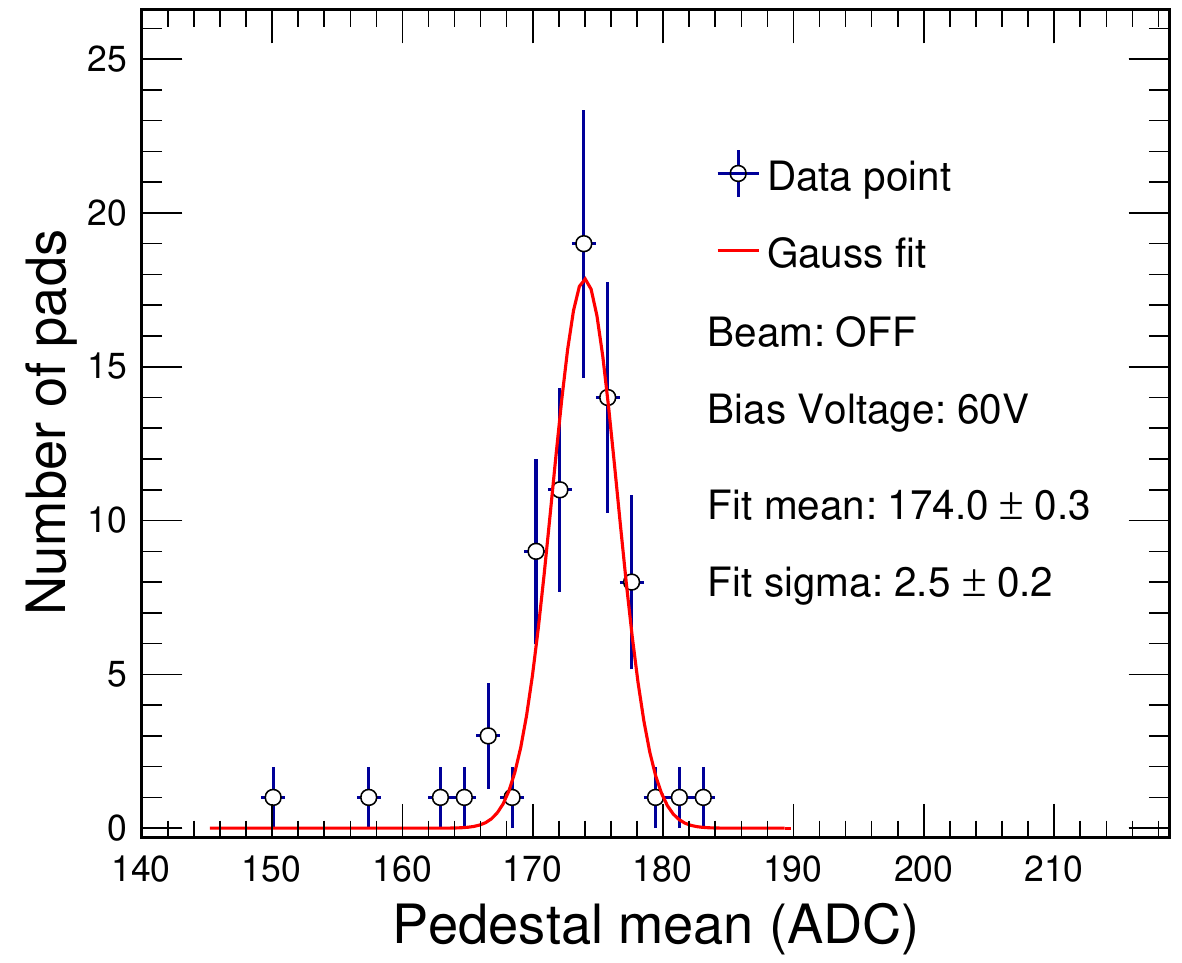}
\caption{Left: An example of the pedestal distribution in a pad out of 72 pads in the detector. Right: Mean value of the pedestal for all 72 pads in the detector. The error bars represent the statistical uncertainties. \label{fig:noise}}
\end{figure}

Background noise is estimated to evaluate the combined noise contributions from the detector and associated electronics. In the absence of the beam, the background data is collected with the detector operated at a reverse bias voltage of 60~V, slightly more than the full depletion voltage of about 50~V. The full depletion voltage is determined from the capacitance vs voltage (CV) measurements. Due to the stochastic or random nature of noise, the background distribution, referred to as a pedestal, is fitted with the Gaussian distribution. The left panel in Figure~\ref{fig:noise} shows the pedestal for a single pad in the detector fitted with a Gaussian function. The right panel in Figure~\ref{fig:noise} shows the pedestal mean values of all 72 pads fitted with the Gaussian function, which peaks around 174~ADC. The data analysis for the electron and pion measurements is done after subtracting the pedestal mean and three times the standard deviation separately for all the pads.

\subsection{Detector response to pion beam} \label{sec: response to pion beam}
\begin{figure}[t]
\centering
\includegraphics[width=0.47\textwidth]{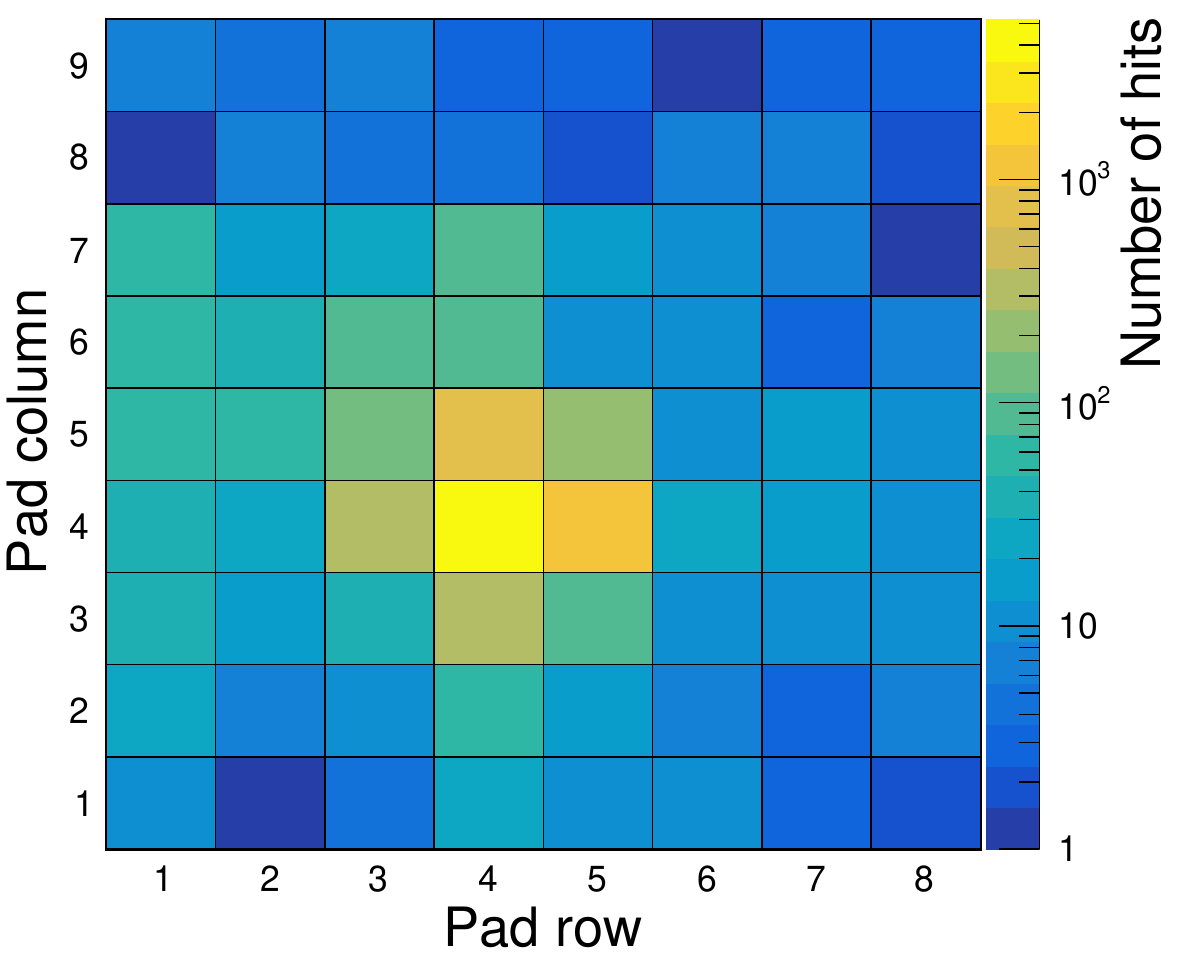}
\qquad
\includegraphics[width=0.47\textwidth]{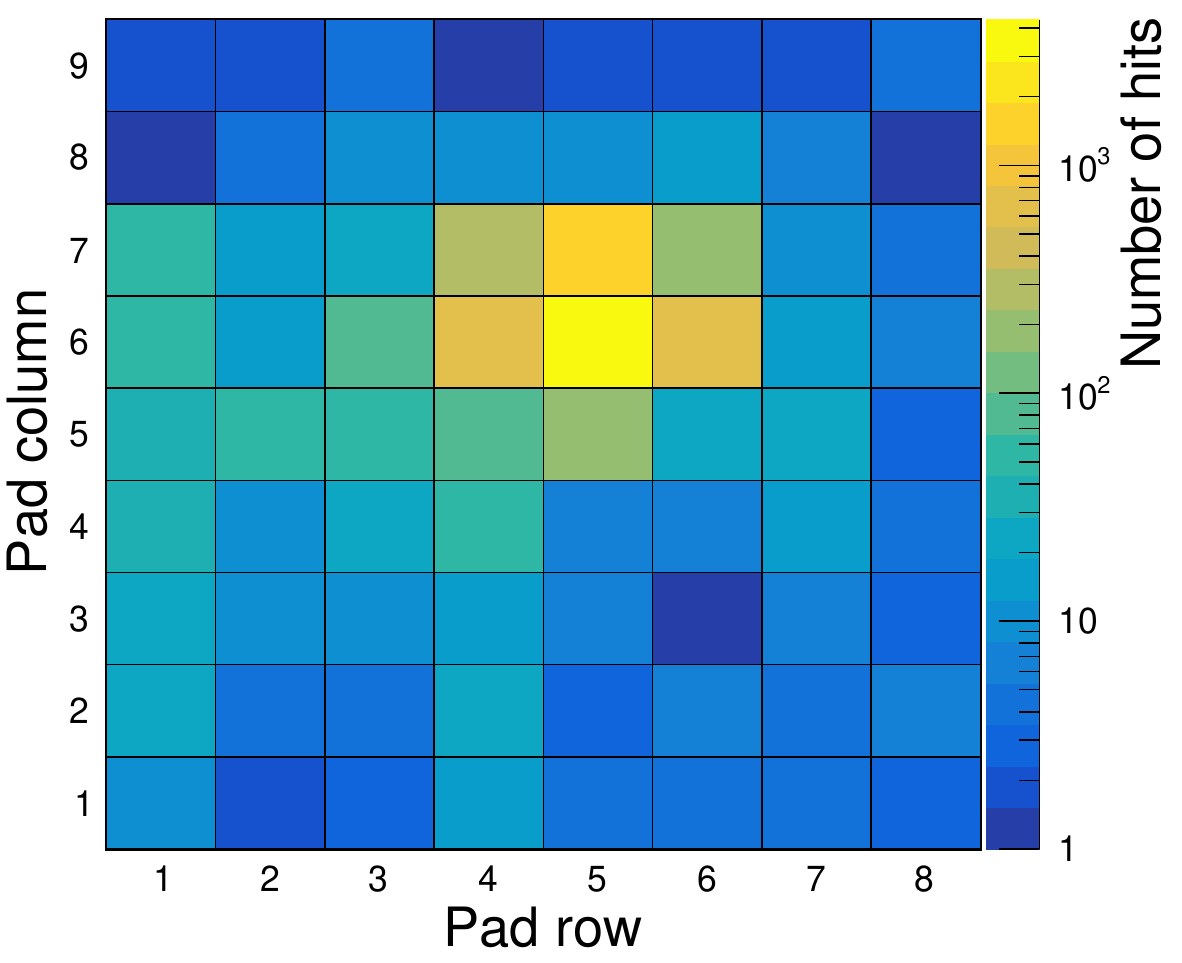}
\caption{Hit map of 10~GeV pion beam focused on two different Si pads during the position scan. The colors reflect the number of hits on the pad. \label{fig:pion_hitmap}}
\end{figure}
\begin{figure}[h]
\centering
\includegraphics[width=.47\textwidth]{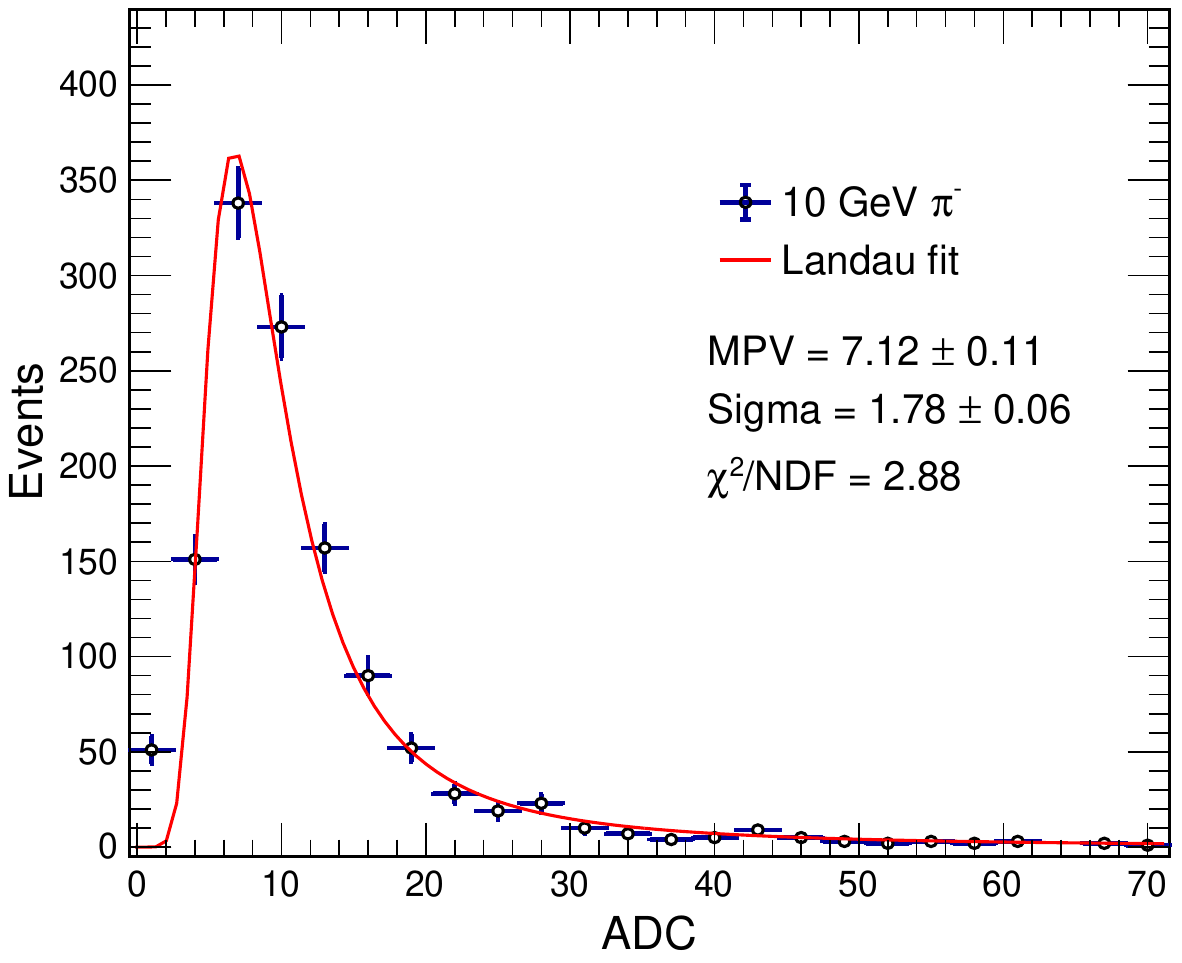}
\qquad
\includegraphics[width=.47\textwidth]{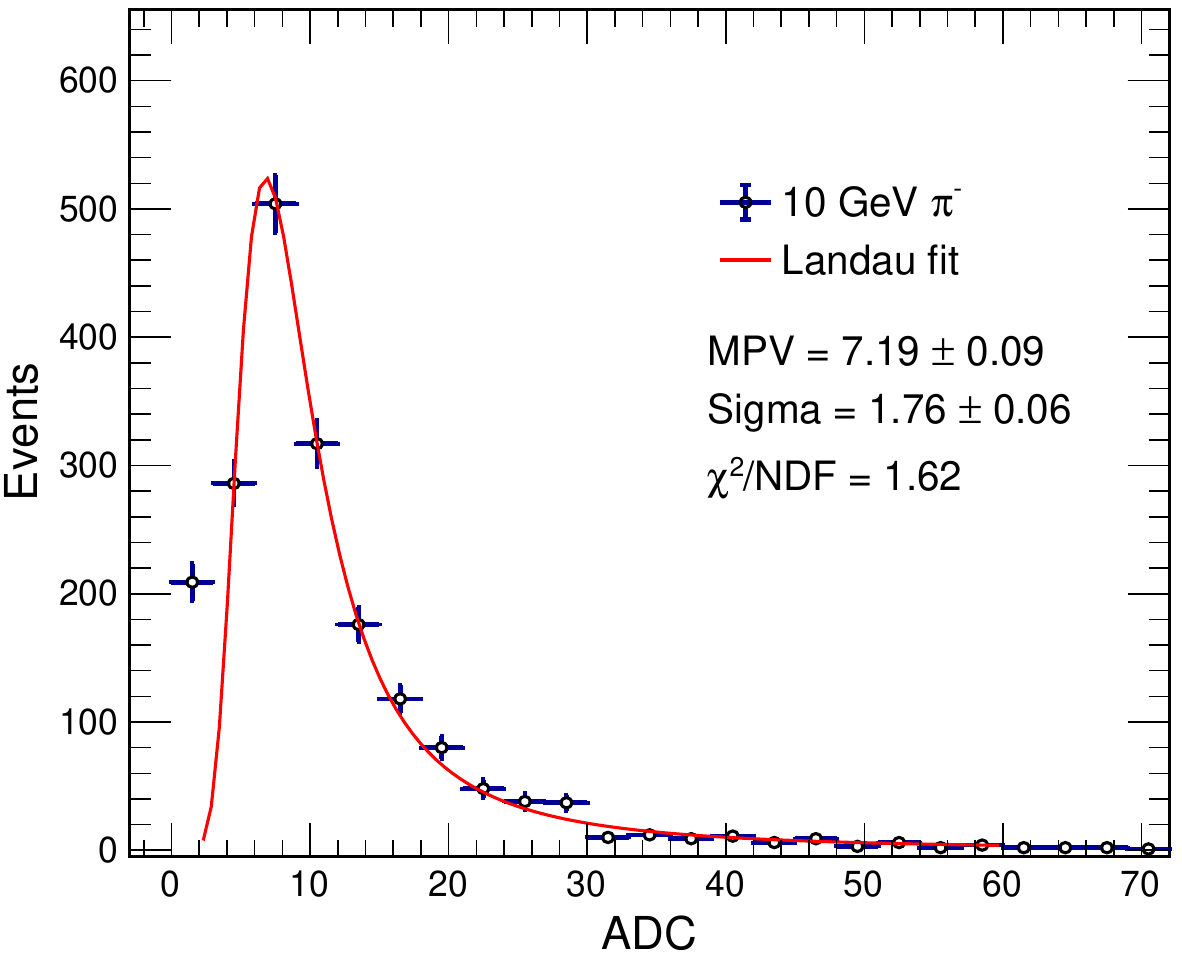}
\caption{Energy deposition of a 10~GeV pion beam focused on two different pads of the $\mathrm{8\times9}$ Si pad array.\label{fig:pion_mip}}
\end{figure}

The term \textit{Minimum Ionizing Particle} (MIP) refers to a particle that passes through matter with the least amount of energy loss. A pion with energy in $\sim$~GeV range acts as a MIP. The Landau distribution characterizes an MIP because it can effectively account for the energy loss patterns of charged particles passing through thin absorbers or gases. The long tail in the Landau distribution accounts for the occasional significant energy loss by the particle from bremsstrahlung, leading to an asymmetric distribution. The position of the Landau peak defines the \textit{Most Probable Value} (MPV) of the energy loss~\cite{LeoBook}. To characterize the response of the silicon detector, it is tested with a 10~GeV pion without any W plates in front of the detector. Figure~\ref{fig:pion_hitmap} shows the 2-D hit maps of the detector. A pad hit is defined as the ADC value above the noise. The spread of hits in the hit map in regions nearby to the beam axis is because the intersection area of two small plastic scintillators ($\sim$~1.44~cm$^2$) is greater than the pad size of 1 cm$^2$. The pion data after the noise subtraction shows the Landau distribution with the MPV around $\sim$7~ADC, as shown in Fig.~\ref{fig:pion_mip} for two different pads in the detector.

\subsection{Detector response to electron beam} \label{sec: response to electron beam}
\begin{figure}[t]
\centering
\includegraphics[width=0.47\textwidth]{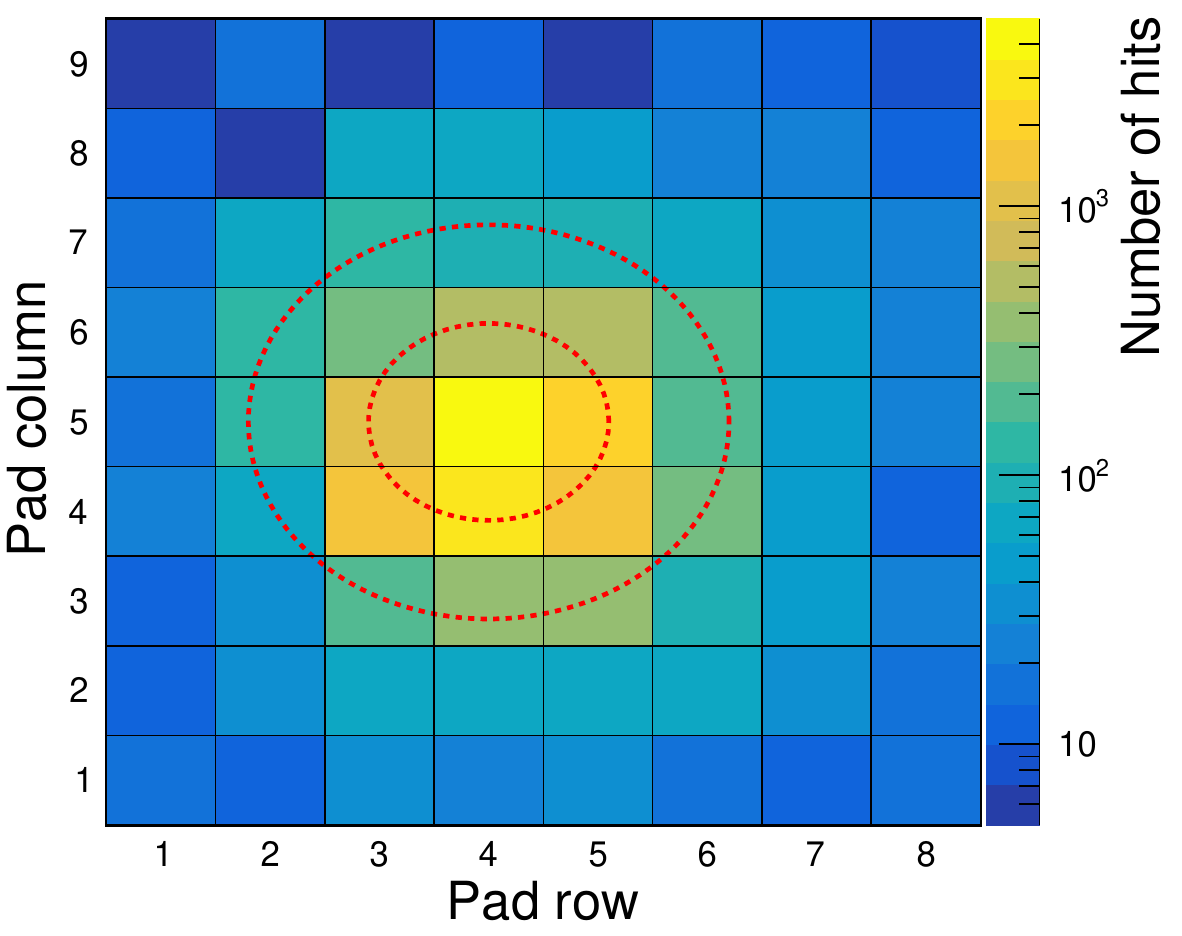}
\qquad
\includegraphics[width=0.47\textwidth]{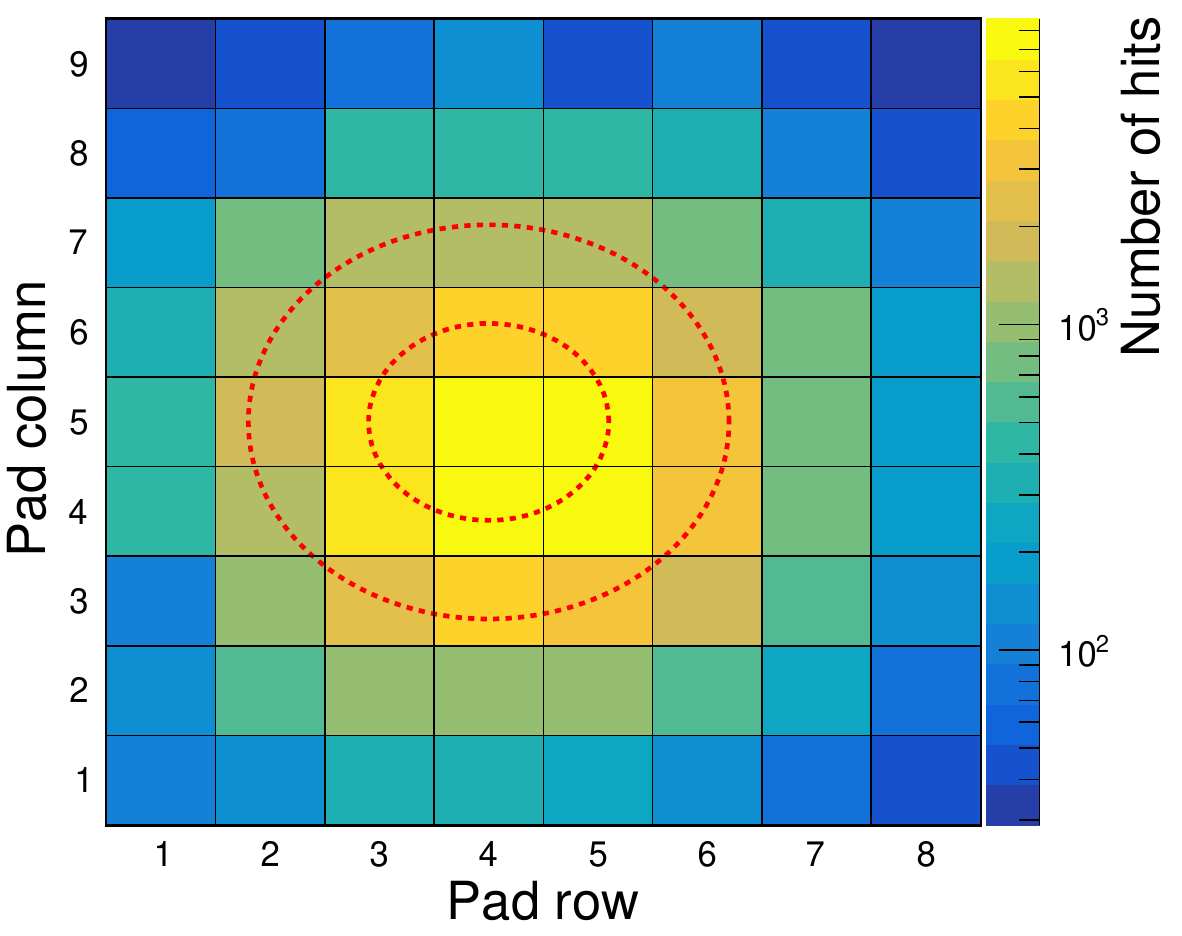}
\caption{$\mathrm{8\times9}$ pad array hit map of 5~GeV electrons across 1~$\mathrm{X_0}$ (left) and 5~$\mathrm{X_0}$ (right). Colors reflect the number of hits in the pad. The inner red circle marks a total of 9 pads, including the pad with max hit and its nearest neighbors, and the outer circle marks the next nearest 16 neighboring pads. The spread in the hits is more for across 5 $\mathrm{X_0}$ than 1~$\mathrm{X_0}$ due to the generation of $\mathrm{e^+e^-}$ shower.} \label{fig:hitmap_electrons}
\end{figure}

In this section, the response of the detector to the electron beam is reported. When electrons interact with matter, they lose energy through different mechanisms depending on their energy and the nature of the detector material. The main energy-loss mechanism is a combined effect of collisional and radiative loss~\cite{LeoBook}. In the collisional loss, the electrons interact with the atoms of the material and transfer energy via ionization or the excitation of atoms. In the radiative loss, the electrons interact with the nucleus of the material via the emission of bremsstrahlung radiation. The ratio of these two mechanisms depends on the atomic number (Z) of the detector material and the energy (E) of the electrons. For a given material, there is a critical energy ($\mathrm{E_{c}}$) at which the collisional and radiative losses become equal. For energies below $\mathrm{E_{c}}$, the collisional loss dominates, while for energies above $\mathrm{E_{c}}$, radiative loss dominates. For tungsten, which has a high atomic number of 74, the critical energy is about 8~MeV. Therefore, for electron beams with energies of the order of GeV, the energy loss is mainly due to radiative loss, which results in the production of electromagnetic showers.\\
The study of electromagnetic showers is performed by varying the total absorber depth in front of the detector. Tungsten absorbers used in the setup have a thickness of about 1~$\mathrm{X_0}$ each. By changing the number of W plates, the shower is studied as a function of $\mathrm{X_0}$.
The 2-D hit map of 5~GeV electrons at 1 and 5 radiation lengths ($\mathrm{X_0}$) is shown in Fig.~\ref{fig:hitmap_electrons}. Across 1~$\mathrm{X_0}$, the percentage of hits in total nine pads around the pad with maximum hits (shown by the inner red circle) is 80.8$\%$ while in the outer 16 pads (shown by the outer red circle) is 12.7$\%$. The remaining hits are outside the outer circle. For 5~$\mathrm{X_0}$, the percentage of hits in the inner nine pads is 54.6$\%$ while in the outer 16 pads is 30.8$\%$. The more dispersion of hits in the lateral dimensions of the detector for 5~$\mathrm{X_0}$ in comparison to 1~$\mathrm{X_0}$ is due to the development of electromagnetic shower in the absorber. This effect is studied in detail in Sec.~\ref{sec: longitudinal shower profile} \& \ref{sec: transverse shower profile}, where the longitudinal and transverse electromagnetic shower profiles of various electron energies are reported. 

\begin{figure}[htbp]
\centering
\includegraphics[width=0.47\textwidth]{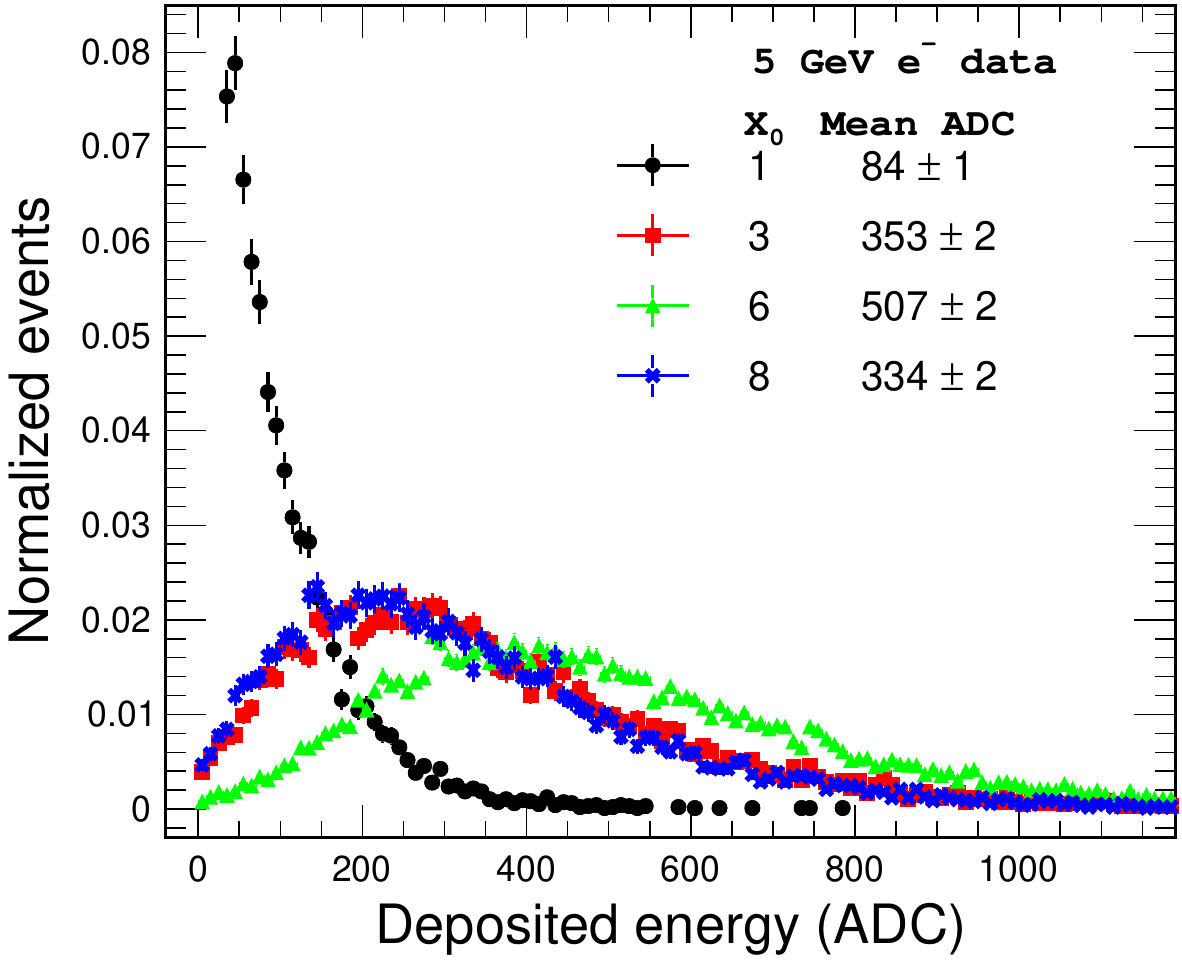}
\qquad
\includegraphics[width=0.47\textwidth]{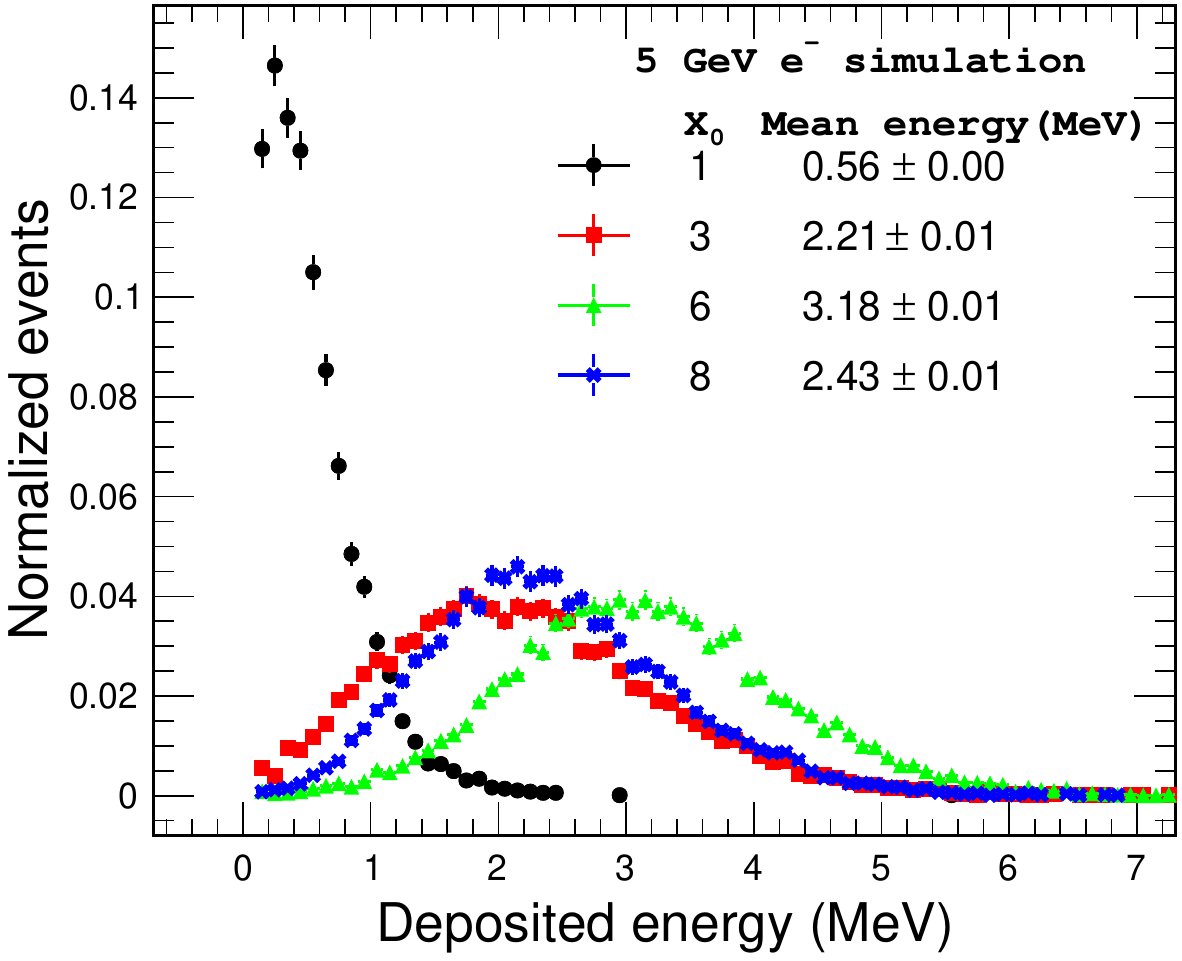}
\caption{Cluster ADC distribution (left) and simulated energy distribution (right) of 5~GeV electrons at various radiation lengths in the detector. \label{fig:cluster ADC distributions data}}
\end{figure}

\subsubsection{Energy deposition in silicon pad array} \label{sec: cluster adc distribution}
The data obtained for electrons with energies from 1~GeV to 5~GeV at different $\mathrm{X_0}$ is analyzed using a clustering algorithm (nearest neighbour algorithm). In this algorithm, the adjacent or connected pads forming a cluster around the pad with maximum hits are kept for the analysis. The clustering algorithm helps to filter out the uncorrelated responses that do not arise from particle interactions with the pads in the detector. The energy deposited by the shower in the detector is obtained by summing the ADC values of the pads in clusters. Figure~\ref{fig:cluster ADC distributions data} (left) shows the cluster ADC distribution and (right) the simulated energy deposition of 5~GeV electrons. The mean of both distributions increases with the $\mathrm{X_0}$ until it reaches a shower maximum and then starts to decrease. The measured energy deposited in data (ADC) is compared with the energy deposited in simulations (MeV) using a calibration factor, which is discussed in the next section.

\begin{figure}[htbp]
\centering
\includegraphics[width=0.9\textwidth]{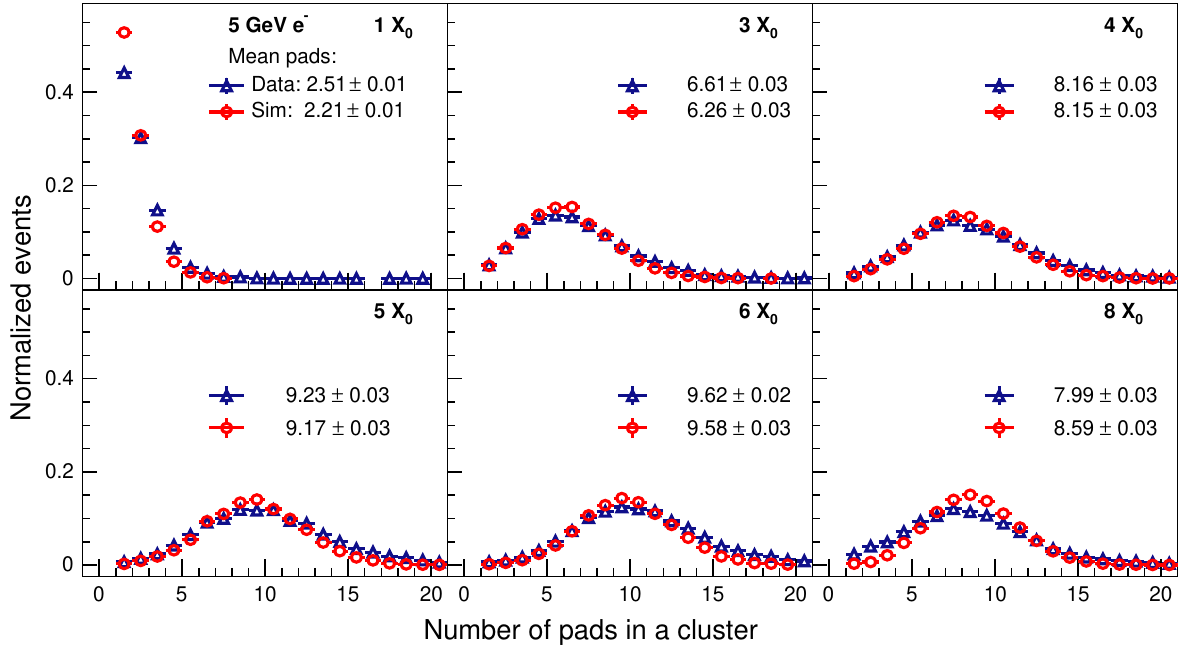}
\caption{Number of pads forming the cluster plotted for several radiation lengths for 5~GeV electron in data (blue) and simulation (red).\label{fig:cluster pads distributions data}}
\end{figure}

The electromagnetic shower is also studied using the cluster pad distribution, which reflects the transverse spread in the shower as a function of $\mathrm{X_0}$. This distribution measures the number of pads in a cluster. Figure~\ref{fig:cluster pads distributions data} shows cluster pad distribution for 5~GeV electrons in the data and simulations. At low $\mathrm{X_0}$ values, the distributions are non-Gaussian and skewed towards the right, indicating that the shower is narrow as it starts to form. At high $\mathrm{X_0}$ values, the distributions become Gaussian and have higher mean values, implying that the shower is more spread out. The data and simulations match well. Section~\ref{sec: longitudinal shower profile} discusses the longitudinal shower profiles in detail.

\subsubsection{Energy calibration} \label{sec: calibration}
\begin{figure}[htbp]
\centering
\includegraphics[width=0.6\textwidth]{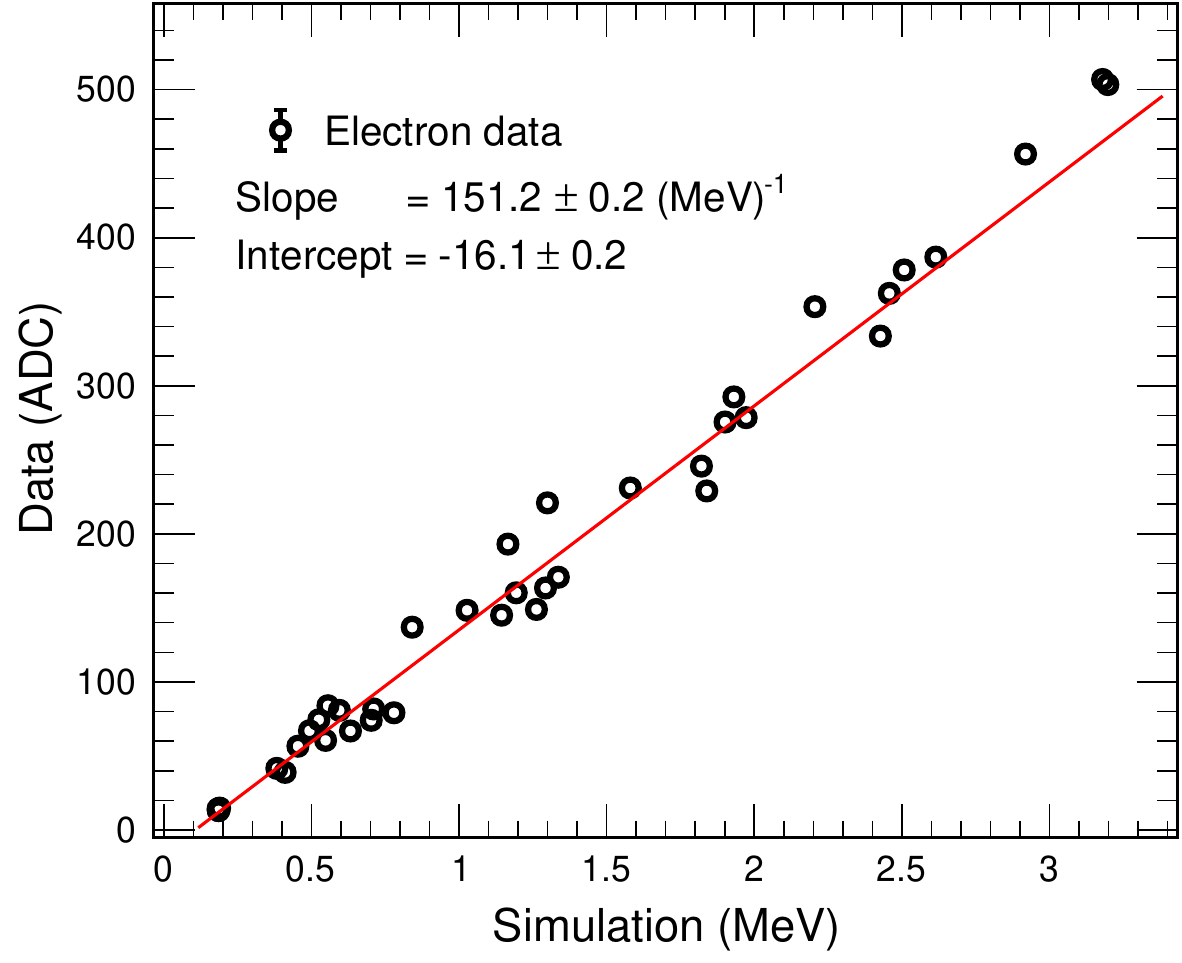}
\caption{Energy deposited by electrons in the detector in experimental data (ADC) against the deposited energy in simulations (MeV) shows a linear trend. Due to large statistics, the error bars are small and are within marker size.\label{fig:calibration}}
\end{figure}

The energy calibration converts the ADC unit to the MeV unit by comparing the experimental data with the simulations. The mean of cluster ADC distribution in data is plotted against the mean of cluster energy distribution in simulations for different electron energies and radiation lengths (see Fig.~\ref{fig:cluster ADC distributions data} for a subset of the data). Figure~\ref{fig:calibration} shows the linear relationship between the two variables. A linear fit is applied to the distribution, and the fit parameters are used for the calibration. The calibrated data is then compared with the simulations in the next section.


\subsubsection{Longitudinal shower profile}\label{sec: longitudinal shower profile}
\begin{figure}[t]
\centering
\includegraphics[width=0.47\textwidth]{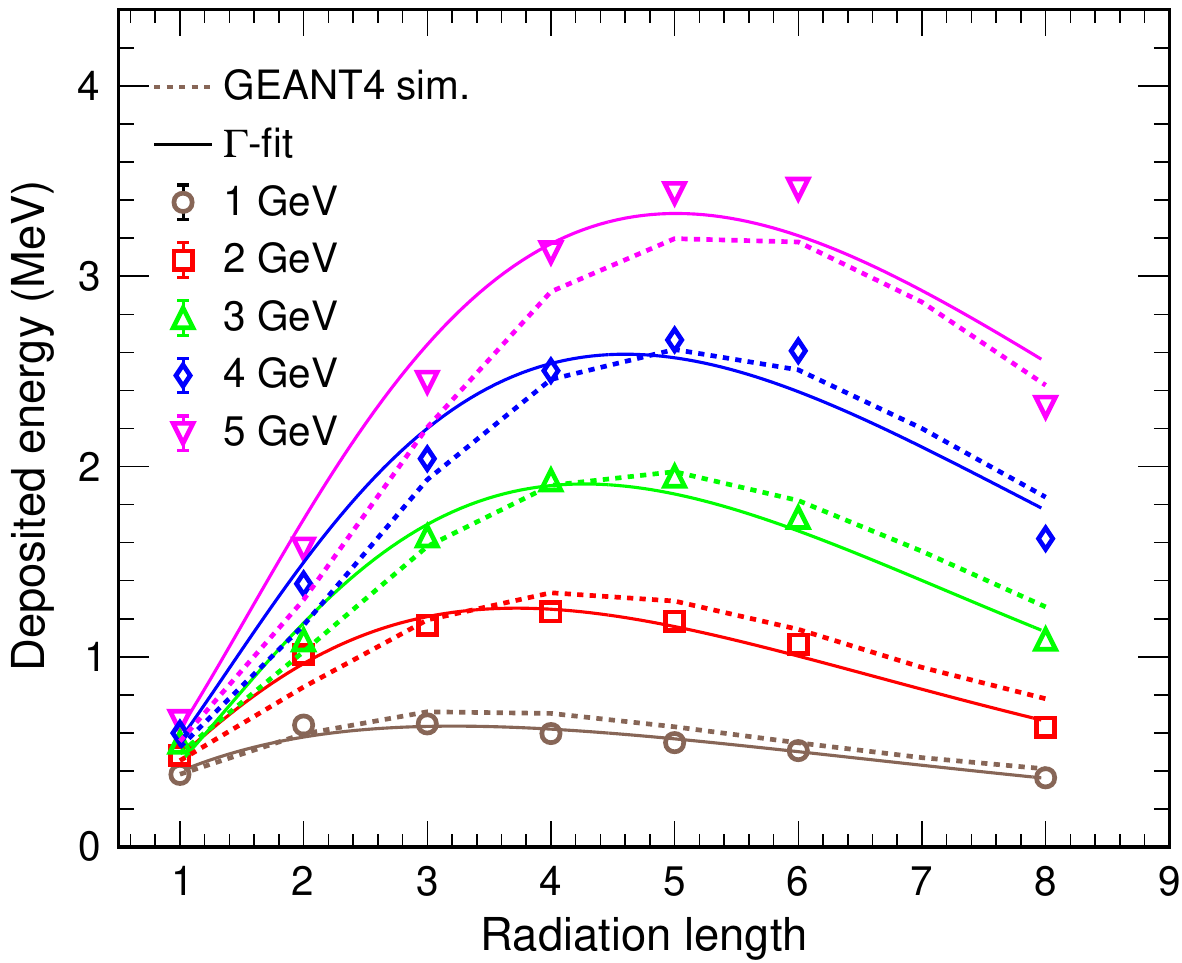}
\qquad
\includegraphics[width=0.47\textwidth]{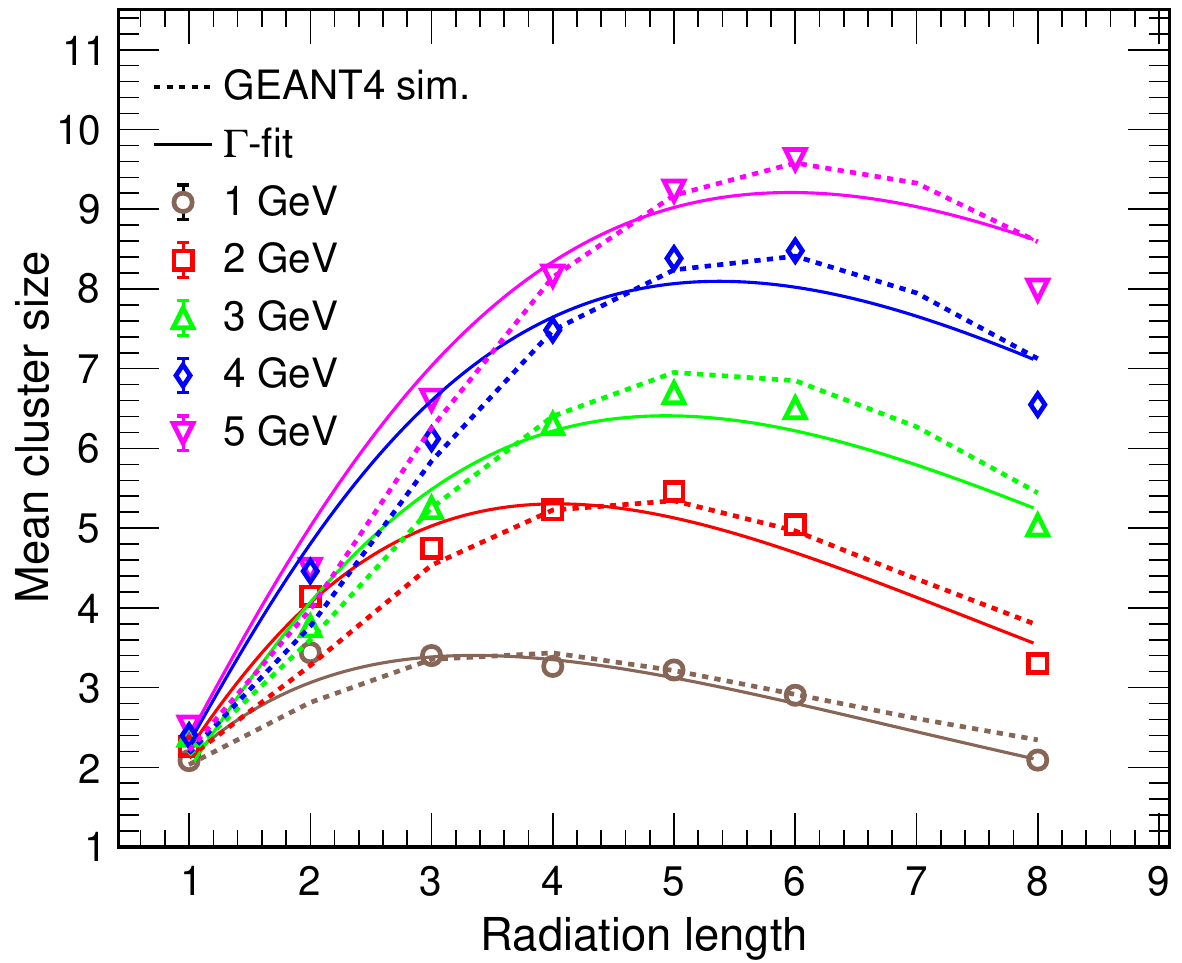}
\caption{Longitudinal shower profile (left) and mean cluster size distribution (right) are plotted for electrons with energies ranging from 1~GeV to 5~GeV, fitted using the $\Gamma$-distribution (solid line) and compared with simulations (dotted line). Due to large statistics, the error bars in the data are small and are within the marker size.\label{fig:shower_profile}}
\end{figure}

The development of an electron shower in the direction of the beam is studied using the longitudinal shower profile and the mean cluster size distribution. The longitudinal shower profile is the average shower energy deposited per radiation length obtained from the mean value of cluster ADC distributions (see Fig.~\ref{fig:cluster ADC distributions data}). Fig.~\ref{fig:shower_profile} (left) shows the longitudinal shower profile for electrons of different energies in data and simulations. The simulations are consistent with the data and show that the deposited energy in the shower profile initially increases with radiation length, which reflects the generation of secondary particles due to the process of pair production and bremsstrahlung. The shower profile then reaches a maximum called shower maximum, and then it starts to decrease due to the fall of energy of cascading particles below the critical energy. The same trend is observed in the mean cluster size distribution, which measures the mean number of pads in a cluster as a function of radiation length. The mean number of pads initially increases with radiation length due to the widening of the shower, reaches a maximum, and then starts to fall gradually due to absorption of low energy particles, reducing the cluster size as shown in the Fig.~\ref{fig:shower_profile} (right).
The longitudinal shower profile and the mean cluster size distribution are characterized by the $\Gamma$-distribution function, which is illustrated in Eq.~\ref{eq:gamma_function}~\cite{LeoBook}. The $\Gamma$-function fits the distributions well, except for the case of zero $\mathrm{X_0}$, where it approaches zero. However, the data has a non-zero value at zero~$\mathrm{X_0}$ due to the minimum deposition of energy and formation of the small clusters by the incoming electrons in the beam.

\begin{equation}
\label{eq:gamma_function}
\begin{aligned}
\frac{dE}{dx} = E_0 b \frac{(bt)^{a-1} e^{-bt}}{\Gamma(a)}
\end{aligned}
\end{equation}
In the equation Eq.~\ref{eq:gamma_function}, the parameter $t$ corresponds to the radiation length, $E_0$ represents the initial energy of the electron, and $a$ and $b$ serve as the fitting parameters that describe the scale and shape of the distribution. The radiation length at which the shower maximum occurs can be determined by taking the derivative of Eq.~\ref{eq:gamma_function} and setting it to zero, leading to $\mathrm{t_{max} = \frac{a-1}{b}}$. The shower maximum obtained using the fitting parameters $a$ and $b$ for electrons with energies ranging from 1~GeV to 5~GeV is within 4~$\mathrm{X_0}$ to 6~$\mathrm{X_0}$, which is in accordance with the data and simulations.

\subsubsection{Transverse shower profile}\label{sec: transverse shower profile}

\begin{figure}[htbp]
\centering
\includegraphics[width=0.6\textwidth]{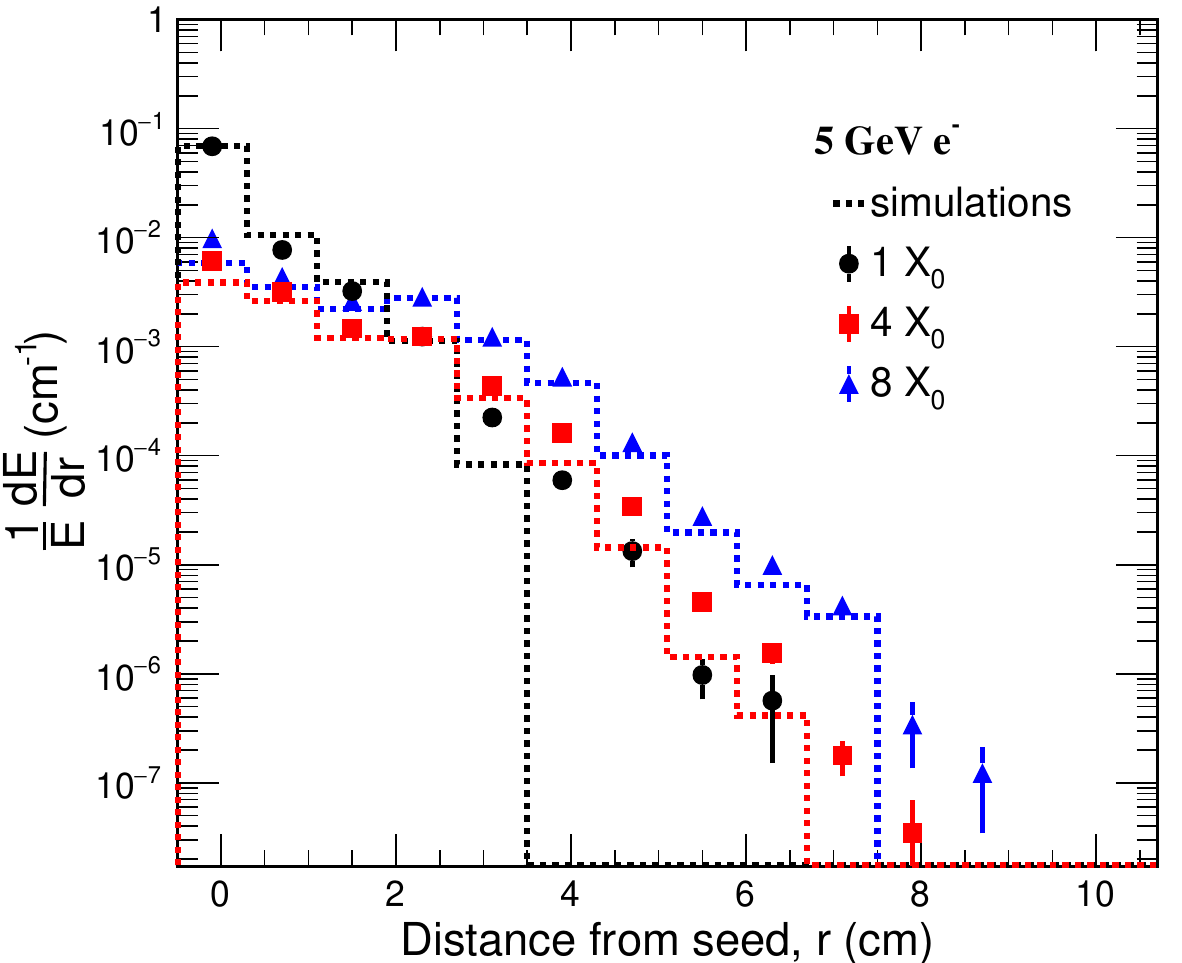}
\caption{Transverse shower profile of 5~GeV electrons at three different depths in the calorimeter: 1~$X_{0}$, 4~$X_{0}$, and 8~$X_{0}$. The horizontal axis represents the distance from the cluster seed ($r$), and the vertical axis (log scale) represents the fraction of energy deposited within radius $r$. \label{fig: transverse profile}}
\end{figure}

The transverse shower profile is another important characteristic of electromagnetic calorimetry. It describes the distribution of energy deposited by the shower particles in the calorimeter in the direction perpendicular to the beam axis. The transverse shower profile is influenced by various effects, such as the non-zero opening angle between the electron and positron pair, which causes them to spread out from the shower axis, the multiple scattering of the electrons, which deflects them from their original direction, and the emission of bremsstrahlung photons at arbitrary angles~\cite{LeoBook}. Fig.~\ref{fig: transverse profile} shows the transverse shower profile, which measures the fraction of energy deposited within a radius $r$ from the cluster seed. It compares the transverse shower profile for 5~GeV electrons for different radiation lengths in data (markers) and simulations (dotted lines). The profile shows that the fraction of energy shared by the pads in the cluster changes as we go to higher radiation lengths. In 8~$\mathrm{X_0}$, the energy shared by pads away from the cluster seed is more in comparison to 4~$\mathrm{X_0}$ and 1~$\mathrm{X_0}$ due to the spread of electromagnetic shower in the lateral dimensions. The transverse shower profile is an important observable that characterizes the performance of the granular detectors. Their study is essential for the design and optimization of the FoCal detector.

\section{Conclusion and Outlook}
The 8x9 n-type Si pad array detectors were successfully designed and fabricated in Bharat Electronics Limited, Bangalore, India. The detectors are read out using HGCROCv2 ASIC, and their in-beam performance test results are reported, revealing a clear MIP spectrum for the pion beam across different pads on the Si pad array. By using the tungsten plates in front of the detector, its longitudinal and transverse electromagnetic shower profiles were studied. The observed shift of shower max towards higher radiation lengths with increased electron energies agrees with Geant4 simulations. In the future, a full-scale prototype of 20~Si-W detector layers is being prepared to study the energy resolution of the sampling calorimeter. The same will be tested with electrons and pions at PS and SPS test beam area, CERN. 


\acknowledgments
The authors would like to thank Mr. Ranajay Laha and Mr. Arijit Das from Bharat Electronics Limited (BEL), Bangalore, for the design, fabrication, and packaging of the silicon pad array detector, Mr. Debasis Barik (Scientific Assistant, CMRP NISER) for help with Mechanical support structure, Mr. Samar Mohan Mohanty (Project associate, CMRP) the ALICE FoCal collaboration and the CERN PS accelerator support team for constant support throughout the project work. We thank the in charge of the T9 facility at CERN, Dipanwita Banerjee, for all the help extended to us. Additionally, the authors would like to thank DAE and DST India for their financial support through the project entitled "Indian participation in the ALICE experiment at CERN," and the work is also partly funded through the J.C. Bose fellowship of DST awarded to BM. We also acknowledge the use of the Garuda HPC facility at the School of Physical Sciences, NISER.


\bibliographystyle{JHEP}
\bibliography{biblio.bib}
\end{document}